\documentclass[twocolumn,pra,superscriptaddress,notitlepage]{revtex4-1}

\usepackage{amsmath}
\usepackage{graphicx} 
\usepackage{here}

\usepackage{amsmath,amssymb,amsthm,mathrsfs,amsfonts,dsfont}
\usepackage{braket}
\usepackage{bm}
\usepackage{enumerate}
\usepackage{color}

\usepackage{algorithm}
\usepackage{algorithmic}
\usepackage{comment}%

\newcommand{\nn}{\notag \\}

\newcommand{\A}{\mathrm{A}}
\newcommand{\B}{\mathrm{B}}
\newcommand{\C}{\mathrm{C}}
\newcommand{\D}{\mathrm{D}}
\newcommand{\tr}{\mathrm{Tr}}%
\newcommand{\hH}{\hat{H}}%
\newcommand{\hU}{\hat{U}}
\newcommand{\hV}{\hat{V}}
\newcommand{\hsig}{\hat{\sigma}}
\newcommand{\hrho}{\hat{\rho}}

\begin{document}


\title{Calculation of Gibbs partition function with imaginary time evolution on near-term quantum computers}


\author{Keisuke Matsumoto  }
\email{1221544@ed.tus.ac.jp, matsumoto.kei@aist.go.jp}
\affiliation{Department of Physics, Tokyo University of Science,
1-3, Kagurazaka, Shinjuku-ku, Tokyo, 162-8601, Japan.}
\affiliation{Research Center for Emerging Computing Technologies, National Institute of Advanced Industrial Science and Technology (AIST), Umezono1-1-1, Tsukuba, Ibaraki 305-8568, Japan.}

\author{Yuta Shingu  }
\affiliation{Department of Physics, Tokyo University of Science,
1-3, Kagurazaka, Shinjuku-ku, Tokyo, 162-8601, Japan.}
\affiliation{Research Center for Emerging Computing Technologies, National Institute of Advanced Industrial Science and Technology (AIST), Umezono1-1-1, Tsukuba, Ibaraki 305-8568, Japan.}

\author{Suguru Endo}
\affiliation{NTT Computer and Data Science Laboratories, NTT Corporation, Musashino, Tokyo 180-8585, Japan.}

 \author{Shiro Kawabata}
 \affiliation{Research Center for Emerging Computing Technologies, National Institute of Advanced Industrial Science and Technology (AIST), Umezono1-1-1, Tsukuba, Ibaraki 305-8568, Japan.}

 \author{\ \ \ \ \ \ \ \ \ \ \ \ \ \ \ \ \ \ \ \ \ \ \ \ \ \ \ \ \ \ \ \ \ \ \ \ \ \ \ \ \ \ \ \ \ \ \ \ \ \ \ \ \ \ \ \ Shohei Watabe}
 \affiliation{Department of Physics, Tokyo University of Science,
1-3, Kagurazaka, Shinjuku-ku, Tokyo, 162-8601, Japan.}

 \author{Tetsuro Nikuni}
  \email{nikuni@rs.kagu.tus.ac.jp}
 \affiliation{Department of Physics, Tokyo University of Science,
1-3, Kagurazaka, Shinjuku-ku, Tokyo, 162-8601, Japan.}

\author{Hideaki Hakoshima}
\affiliation{Research Center for Emerging Computing Technologies, National Institute of Advanced Industrial Science and Technology (AIST), Umezono1-1-1, Tsukuba, Ibaraki 305-8568, Japan.}

\author{Yuichiro Matsuzaki 
}
\email{matsuzaki.yuichiro@aist.go.jp}  
\affiliation{Research Center for Emerging Computing Technologies, National Institute of Advanced Industrial Science and Technology (AIST), Umezono1-1-1, Tsukuba, Ibaraki 305-8568, Japan.}


\begin{abstract}
The Gibbs partition function is an important quantity in describing statistical properties of a system in thermodynamic equilibrium. There are several proposals to calculate the partition functions on near-team quantum computers. However, the existing schemes require many copies of the Gibbs states to perform an extrapolation for the calculation of the partition function, and these could be costly performed on the near-term quantum computers. Here, we propose an efficient scheme to calculate the Gibbs function with the imaginary time evolution.
To calculate the Gibbs function of $N$ qubits, only $2N$ qubits are required in our scheme.
After preparing Gibbs states with different temperatures
by using the imaginary time evolution, we measure the overlap between them on a quantum circuit, and this allows us to calculate the Gibbs partition function. 
\end{abstract}

\maketitle

\section{Introduction}
In equilibrium statistical mechanics~\cite{Feynman1998}, the Gibbs partition function $Z$ is an important quantity.
From the partition function, one can calculate the free energy $F=-k_{\rm{B}}T {\rm {ln}} Z$ where $k_{\rm{B}}$ denotes the Boltzman constant and $T$ the temperature, and the free energy provides useful information about thermodynamic properties of the system. 
However, generally speaking, it is difficult to calculate the partition function for a given microscopic Hamiltonian composed of a large number of qubits. When we diagonalize the Hamiltonian, this is usually not tractable for classical computers when the number of the qubits is large.

Many efforts have been made to develop quantum algorithms for the Noisy Intermediate-Scale Quantum (NISQ) devices~\cite{preskill2018quantum}.
Such a NISQ device could contain tens to thousands of qubits, and the error rate would be around $10^{-3}$~\cite{endo2021hybrid} .
Variational quantum algorithms (VQAs)~\cite{peruzzo2014variational,kandala2017hardware,moll2018quantum,mcclean2016theory,farhi2014quantum,li2017efficient,yuan2019theory} have been considered as one of the most promising applications of the NISQ devices.
There are several VQAs, such as for quantum chemistry, machine learning, and finance~\cite{mcardle2020quantum,cao2019quantum,mitarai2018quantum,benedetti2019generative}.
Among other VQAs, variational quantum simulation (VQS) allows us to simulate imaginary time evolution of quantum systems~\cite{mcardle2019variational}.
This approach is known to be useful to estimate the energy of the ground state of the Hamiltonian.
Moreover, via the imaginary time evolution of a total system composed of the original system and ancillary system, one can prepare a specifc state called the thermofield double (TFD) states, which can be used to obtain a Gibbs state $\hat{\rho}_{\rm{th}}$ of the original system by tracing out the ancillary one~\cite{yuan2019theory}. Other methods to prepare the Gibbs state have been proposed in Refs.~\cite{wu2019variational,chowdhury2020variational,wang2020variational,tan2020quantum,motta2020determining,francis2020body,harsha2020wave,cohn2020minimal,shingu2021boltzmann}.

There are several existing methods to calculate the free energy~\cite{wu2019variational,chowdhury2020variational,wang2020variational,tan2020quantum,bassman2021computing}.
For example, one can calculate the free energy as $F=E-TS$, where
 $E={\rm{Tr}}[\hat{\rho}_{\rm{th}}\hat{H}]$ denotes the internal energy
 and 
$S=-{\rm{Tr}}[\hat{\rho}_{\rm{th}}{\rm{ln}}\hat{\rho}_{\rm{th}}
]$ denotes the von Neumann entropy.
One can calculate the internal energy on a quantum computer by choosing the Hamiltonian as an observable.
On the other hand, the direct calculation of the von Neumann entropy is difficult. The previous study~\cite{wu2019variational} proposed a method using an extrapolation of R{\'e}nyi entropy;
A R{\'e}nyi entropy is defined as
\begin{equation}
S_p(\hat{\rho}_{\rm{th}}) =\frac{1}{1-p} {\rm{ln}} {\rm{Tr}}[
\hat{\rho}_{\rm{th}}^p],
\end{equation}
and it is known that requires the von Neumann entropy by taking a limit of $p\rightarrow 1$~\cite{zyczkowski2003renyi,fannes2012connecting,johri2017entanglement}.
In the conventional approaches, the higher order of the R{\'e}nyi entropy are cauclated on a quantum computer, and the von Neumann entropy is estimated by extrapolating to $p=1$.
However, in order to calculate the $p$-th order R{\'e}nyi entropy, the necessary number of qubits is $\Theta(p N)$, and thus one needs need to increase the number of the qubits to improve the accuracy of the extrapolation. This might cause difficulty in implementing this scheme on a near-term quantum computer that has the limited number of the qubits.

In this paper, we propose a scheme to calculate the partition function with a smaller number of qubits by using the variational imaginery time evolution. 
Nothing that the normalization factor of the wave function during the imaginary time evolution is the partition function of the corresponding Hamiltonian, we develop a systematic method to calculate the normalization factor from an overlap between the wave functions during the imaginary time evolution.
As long as the variational imaginary time evolution is exact, we can directly calculate the partition function 
without the extrapolation method. 
Since our scheme requires only $2N$ qubits to calculate the partition function of $N$ qubits,
the necessary number of the qubits is much smaller than that with
the conventional schemes.
To illustrate the performance of our scheme, we adopt the Heisenberg model composed of two qubits, and show that our method can accurately calculate the partition function of the Heisenberg model with our method.

The paper is organized as follows. In Sec. II, we review the imaginary time evolution by using the VQA. In Sec. III, we describe our scheme to calculate the partition function.
In Sec. IV, we show the results of the numerical simulation to calculate the partition function of the Heisenberg model.
In
Sec. V, we conclude our discussion.
Throughout this paper, we set
$\hbar=k_{\B}=1$.

\section{
VARIATIONAL IMAGINARY TIME
EVOLUTION
}
In this section, we review the variational imaginary time evolution with the NISQ devices~\cite{mcardle2019variational}.
For a given Hamiltonian $\hH$,
the imaginary time evolution is described as
\begin{equation}\label{scheqimagynary}
	-\frac{\partial}{\partial \tau}\ket{\psi(\tau)}=(\hH-E({\tau}))\ket{\psi(\tau)},
\end{equation}
where
\begin{equation}
    E({\tau})=\braket{\psi(\tau)|\hH|\psi(\tau)}.
\end{equation}
Given an initial state $\ket{\psi(0)}$, the state after a time $\tau $
is given by
\begin{equation}\label{imgschsol}
	\ket{\psi(\tau)}=\dfrac{1}{\sqrt{A(\tau)}}\exp(-\hH \tau)\ket{\psi(0)},
\end{equation}
where $A(\tau)=\braket{\psi(0)|e^{-2\hH\tau}|\psi(0)}$ denotes a normalisation factor.

The non-unitary imaginary time evolution operator $\exp(-\hH\tau)$ cannot
be directly represented by a quantum circuit.
Instead, we adopt a parameterized trial wave function
$\ket{\phi(\vec{\theta}(\tau))}=\hat{V}(\vec{\theta}(\tau))\ket{\phi(0)}$ 
where 
$\hat{V}(\vec{\theta}(\tau))=\hU_d(\theta_d(\tau))\cdots\hU_1(\theta_1(\tau))$ denotes a unitary operator with $d$ parameters, $\{\hU_i(\theta_i(\tau))\}_{i=1}^d$
denotes a set of parametrized gate in a variational quantum circuit, $\{\theta_i \}_{i=1}^d$ denotes a set of the parameters, and $\ket{\phi(0)}$ denotes the initial state, which is chosen to be equal to $\ket{\psi(0)}$.

Then, for the imaginary time evolution, we need to calculate the evolution of the parameters. 
For this purpose,
we adopt the McLachlan's variational principle~\cite{mclachlan1964variational}.
Let us consider a distance between the exact dynamics and that of the parametrized wave function defined by
 \begin{align}
 	J=\lVert({\partial}/{\partial \tau} + \hH-E(\vec{\theta}(\tau)))\ket{\phi(\vec{\theta}(\tau))}\rVert ^2,
 \end{align}
where
 \begin{equation}
    E(\vec{\theta}(\tau))=\braket{\phi(\vec{\theta}(\tau))|\hH|\phi(\vec{\theta}(\tau))}.
\end{equation}
Following the McLachlan's variational principle, we minimize the distance$J$ as
\begin{equation}
    \delta \|({\partial}/{\partial \tau} + H-E(\vec{\theta}(\tau)))\ket{\phi(\vec\theta(\tau))}\|=0
\end{equation}
under the constraint $\|\ket{\phi(\vec{\theta}(\tau))}\|=1$. 
We obtain the differential equations for $\vec{\theta}(\tau)$
describing the imaginary time evolution for $\vec{\theta}(\tau)$:
\begin{equation}\label{differential equation}
    \sum_{q=1}^{d}M_{pq}\frac{\partial{\theta}_q}{\partial \tau}=C_p,
\end{equation}
where,
\begin{equation}
\label{Mmatrix}
	M_{pq}
	= \mathrm{Re}\left(\frac{\partial \bra{\phi(\vec{\theta}(\tau))}}{\partial \theta_p}\frac{\partial \ket{\phi(\vec{\theta}(\tau))}}{\partial \theta_q}\right),
\end{equation}
and
\begin{equation}
\label{Cvector}
	C_p
	= -\mathrm{Re}\left(\frac{\partial \bra{\phi(\vec{\theta}(\tau))}}{\partial \theta_p} \hH\ket{\phi(\vec{\theta}(\tau))}\right),
\end{equation}
We describe Eq. \eqref{differential equation} as
$\dot{\vec{\theta}}(\tau)=M^{-1}(\tau)\vec{C}(\tau)$, and solve this for a small time interval $\Delta\tau$ as
\begin{equation}\label{parameter update}
    \vec\theta(\tau+\Delta\tau)
	\simeq\vec\theta(\tau)+\dot{\vec\theta}(\tau)\Delta\tau
	=\vec\theta(\tau)+M^{-1}(\tau)\vec{C}(\tau)\Delta\tau.
\end{equation}
In the case of our interest the derivative of
parametrized gates can be represented as
\begin{equation}\label{Eq:derivativeU}
\frac{\partial \hU_{p }(\theta_p)}{\partial \theta_p} = \sum_k f_{k,p}\hU_{p}(\theta_p)\hat{u}_{k,p}, \end{equation}
where $\hat{u}_{k,p}$ denotes a unitary operator and $f_{k,p}$ denotes a scalar number. For example, if the $p$-th unitary operator $\hU_p(\theta_p)$ is a single qubit rotation described as $R_y(\theta_p)=\exp({-i\theta_p\hsig_{y}/2})$, its derivative is given by ${\partial R_y(\theta_p)}/{\theta_p}=-i/2\hsig_y R_y(\theta_p)$. In this case, by choosing $f_{k=1,p}=-i/2$ and $\hat{u}_{k=1,p}=\hsig_{y}$, we can express ${\partial R_y(\theta_p)}/{\partial\theta_p}$ as Eq. \eqref{Eq:derivativeU}. 
Also,
if the $p$-th unitary operator $\hU_p(\theta_p)$ is a controlled rotation $CR_y(\theta_p)=\ket{0}\bra{0}\otimes\hat{I}+\ket{1}\bra{1}\otimes R_y(\theta_p)$, we obtain
\begin{align}
    \frac{\partial CR_y(\theta_p)}{\partial\theta_p}
    &=-\frac{i}{2}\ket{1}\bra{1}\otimes\{R_y(\theta_p)\hsig_y\}\\[10pt]
    &=-\frac{i}{4}CR_y(\theta_p)\{\hat{I}\otimes\hsig_y\}\nn[5pt]
    &\hspace{10pt}+\frac{i}{4}CR_y(\theta_p)\{\hsig_z\otimes\hsig_y\}.
\end{align}
In this case, by choosing $f_{k=1,p}=-i/4$, $f_{k=2,p}=i/4$, $\hat{u}_{k=1,p}=\hat{I}\otimes\hsig_y$, $\hat{u}_{k=2,p}=\hsig_z\otimes\hsig_y$, we can describe ${\partial CR_y(\theta_p)}/{\partial\theta_p}$ as Eq. \eqref{Eq:derivativeU}.
 Thus, the derivative of the parametrized state $\ket{\phi(\vec{\theta}(\tau))}$ is given as
\begin{equation}\label{Eq:derivativeket}
    \frac{\partial \ket{\phi(\vec{\theta}(\tau))}}{\partial \theta_p}
	= \sum_k f_{k,p} \tilde{V}_{k,p}\ket{\phi(0)},
\end{equation}
and, equivalently,
\begin{equation}
	\label{Eq:derivativebra}
	\frac{\partial \bra{\phi(\vec{\theta}(\tau))}}{\partial \theta_p}
	= \sum_k f^\ast_{k,p} \bra{\phi(0)}\tilde{V}^\dag_{k,p},
\end{equation}
where
\begin{align}
    \tilde{V}_{k,p} &= \hU_{d}(\theta_d)\dots \hU_{i+1}(\theta_{p+1})\hU_{p}(\theta_i)\hat{u}_{k,p}\dots \hU_{1}(\theta_1),\\[10pt]
    \tilde{V}^\dag_{k,p} &= \hU_{1}^\dag(\theta_1)\dots \hU_{p-1}^\dag(\theta_{p-1})
	\hat{u}_{k,p}^\dag \hU_{p}^\dag(\theta_p) \dots \hU_{d}^\dag(\theta_d).
\end{align}
Let us
assume that the $N$-qubit
Hamiltonian $\hH$ can be described as $\hH=\sum_l\lambda_l\hat{P}_l$, where $\lambda_l=\frac{1}{2^N}\tr[\hat{P}_l\hH]$ is a real value and $\hat{P}_l$ is a tensor product of Pauli operators.
Then, from Eqs.\eqref{Mmatrix}, \eqref{Cvector}, \eqref{Eq:derivativeket}, \eqref{Eq:derivativebra}, the coefficients $M_{pq}$ and $C_q$ are given by
\begin{align}
		M_{pq}
		&=\sum_{k,l}\mathrm{Re}\left(f_{k,p}^*f_{l,q}\bra{\phi(\vec{\theta}(0))} \tilde{V}_{k,p}^\dag \tilde{V}_{l,q}\ket{\phi(\vec{\theta}(0))}\right),\\[10pt]
		C_p
		&=\sum_{k,l}\mathrm{Re}\left(f_{k,p}^*\lambda_l\bra{\phi(\vec{\theta}(0))}\tilde{V}_{k,p}^\dag \hat{P}_l \hat{V}\ket{\phi(\vec{\theta}(0))}\right).
\end{align}
It is known that
the can be efficiently calculated these values by using quanutm circuits on the NISQ devices~\cite{mcardle2019variational}. 

\section{Calculating the partition function with the imaginary time evolution}\label{theory}
\subsection{Partition function-normalizatoin factor relation}
We describe our
scheme
to calculate the partition function from the imaginary time evolution with NISQ devices.
We assume $\ket{\psi (\tau)}\simeq \ket{\phi(\vec{\theta}(\tau))}$ where $\ket{\psi (\tau)}$ is the solution of the Eq. \eqref{scheqimagynary} and $\ket{\phi(\vec{\theta}(\tau))}$ is the parametrized wave function with parameters calculated by Eq. \eqref{differential equation}.

Supposing that a system A is composed of $N$ qubits, we aim to calculate the partition function of the system A. Additionally, we consider another system B
composed of $N$ qubits. We consider a total Hamiltonian 
$\hH_{\rm{total}}
=\hH\otimes\hat{I}_{\rm{B}}$ where $\hH$ denotes the Hamiltonian acting on the system A and $\hat{I}_{\rm{B}}$ denotes the identity operator acting on the system B.
It is known that, for a given Hamiltonian $\hH$ composed of $N$ qubits, we can prepare the Gibbs state 
by the imaginary time evolution with the total Hamiltonian $\hH_{\rm{total}}
=\hH\otimes\hat{I}_{\rm{B}}$
~\cite{yuan2019theory}. Choosing the initial state as a maximally entangled state
\begin{equation}
    \ket{\phi(0)}=({1}/{\sqrt{2^N}})\displaystyle\sum_{i=0}^{2^N-1}\ket{i}_{\rm{A}}\ket{i}_{\rm{B}},
\end{equation}
where $\ket{i}_{\rm{A},\rm{B}}$ is the computational basis of the system A and B, we obtain the Gibbs state by tracing out the system B after the imaginary time evolution.
Since we adopt this approach for the calculation of the partition function in our scheme,
we set
the initial parameters $\vec\theta(0)$ to prepare the maximally entangled state as
\begin{equation}\label{Eq:initial state}
    \ket{\phi(\vec\theta(0))}=\hV(\vec\theta(0))\ket{0}^{\otimes 2N}
    =\frac{1}{\sqrt{2^N}}\displaystyle\sum_{i=0}^{2^N-1}\ket{i}_{\A}\ket{i}_{\B},
\end{equation}
where $\ket{0}$ denotes an eigenstate 
of the Pauli operator $\hsig _z$ with an eigenvalue $1$.
The key point of our scheme is that,
when we perform the imaginary time evolution on the initial state~\eqref{Eq:initial state} with the Hamiltonian, the normalization factor $A(\tau)$ of the wave functions is given as
\begin{align}\label{Eq:nf}
    A(\tau)&=\braket{\psi(0)|e^{-2\hH_{\A}\tau}|\psi(0)}\nn[5pt]
            &=\frac{1}{2^N}\sum_{i,j=0}^{2^N-1}
            \braket{i|e^{-2\hH_{\A}\tau}|j}_{\A}\delta_{i,j}=Z(2\tau)/2^N,
\end{align}
where $Z(2\tau)$ is the partition function of the Hamiltonian.
We can rewtite Eq.\eqref{Eq:nf} as
\begin{equation}
    Z(2\tau)=\tr_{\rm{A}}[e^{-2\hH_{\A}\tau}]=\sum_{i=0}^{2^N-1}\braket{i|e^{-2\hH_{\A}\tau}|i}_{\A}.
\end{equation}
Furthermore, by replacing $2\tau$ with the inverse temperature $\beta=1/T$, we obtain \begin{equation}
    Z(\beta)=2^NA(\beta/2),
\end{equation}
which means that the partition function can be calculated once the
normalization factor is obtained.
However, there was no known way to directly measure the normalization factor of the variational imaginary time evolution.

\subsection{Recurrence Formula Method (RFM)}
 In this section, we develop the Recurrence Formula Method (RFM) to calculate the normalization factor from an overlap between the wave functions during the imaginary time evolution.
Importantly, we can measure the overlap
$D(\tau,\tau^\prime)=|\braket{
\phi(\vec{\theta}(\tau))|\phi(\vec{\theta}(\tau^\prime))}|^2$
between $\ket{\phi(\vec{\theta}(\tau))}$ and $\ket{\phi(\vec{\theta}(\tau^\prime))}$ 
on a quantum computer,
because this is a
probability to obtain $+1$ by measuring
$\hsig_z$ for every qubit
when we prepare a state of
$\hat{V}^\dag(\tau')\hat{V}(\tau)
\ket{0}^{\otimes 2N} $.
 The overlap with the normalization factor can be rewritten
 as
\begin{align}\label{Def:overlap}
	D(\tau,\tau^\prime)&=
	\frac{A^2((\tau+\tau^\prime)/2)}{A(\tau)A(\tau^\prime)}.
\end{align}

 We also transform the Eq.~\eqref{Def:overlap} to obtain 
 \begin{equation}\label{nomalA}
    A(\tau)=\frac{A^2((\tau+\tau^\prime)/2)}{D(\tau,\tau^\prime)A(\tau^\prime)}.
\end{equation}
By setting
$\tau=n\Delta\tau$ and $\tau^\prime=(n-2)\Delta\tau$ for $(n\geq2)$, we obtain
\begin{equation}
    A(n\Delta\tau)=\frac{A^2((n-1)\Delta\tau)}{D(n\Delta\tau,(n-2)\Delta\tau)A((n-2)\Delta\tau)}.
\end{equation}
Thus, 
$A(n\Delta\tau)$ can be sequentially calculated
as follows:
\begin{align}
	\frac{A(2\Delta\tau)}{A(\Delta\tau)}&
	=\frac{A(\Delta\tau)}{D(2\Delta\tau,0)A(0)},\nn[10pt]
	\frac{A(3\Delta\tau)}{A(2\Delta\tau)}
	&=\frac{A(2\Delta\tau)}{D(3\Delta\tau,\Delta\tau)A(\Delta\tau)},\nn[10pt]
	&\ \ \vdots\notag\\
	\frac{A(n\Delta\tau)}{A((n-1)\Delta\tau)}
	&=\frac{A((n-1)\Delta\tau)}{D(n\Delta\tau,(n-2)\Delta\tau)A((n-2)\Delta\tau)}.\notag
\end{align}
Summarizing these equations, we obtain the following equation:
\begin{equation}\label{summarizeeq}
    A(n\Delta\tau)
	=\frac{A((n-1)\Delta\tau)
	A(\Delta\tau)}{
	D(n\Delta\tau,(n-2)\Delta\tau)
	\cdots D(4\Delta\tau,2\Delta\tau)
	D(2\Delta\tau,0)}.
\end{equation}
This means that, if $A(\Delta\tau)$ is given,  we can calculate $A(n\Delta\tau)$ for any $n$ by measuring the overlap by using a quantum circuit.
For a sufficiently small $\delta \tau$, we can approximate $A(\Delta\tau)$ by using the Taylor expansion
\begin{align}
    A(\Delta\tau)&=\braket{\phi(0)|e^{-2\hH \Delta \tau}|\phi(0)}\nn[10pt]
    &\simeq1-2\Delta \tau\braket{\phi(0)|\hH|\phi(0)}+\cdots,
\end{align}
which can be approximately calculated by a classical computer. 
We call this method the Recurrence Formula Method (RFM).

However, 
as the temperature of interest decreases(i.e., for large $n$),
the necessary number of the overlaps to be measured by the quantum circuit increases
for the calculations of the denominator of the Eq. (\ref{summarizeeq}).
If each experimentally measured overlap is slightly different from the true value, the error accumulates, and it will be difficult to obtain an accurate value of the partition function. 
Therefore, this approach is considered to be effective in determining the partition function at relatively high temperatures.


\subsection{Reversed Overlap Method (ROM)}
 In this subsection, we propose
 an alternative approach, which we call the Reversed Overlap Method (ROM), to effectively calculate the partition function at low temperatures.
We define $\{\ket{\phi_i}\}_{i=0}^{2^N-1}$ ($\{E_i\}_{i=0}^{2^N-1}$)
as the energy eigenstates (eigenvalues)
of the 
 Hamiltonian $\hH$.
 Here, without loss of generality, we can assume that $E_i\leq E_{i+1}$, $(i=0,1,\cdots, 2^N-2)$.
With the above definitions, obtain
 \begin{align}\label{Eq:expectation basis}
	\braket{\phi(0)|e^{-\hH(\tau+\tau^\prime)}|\phi(0)}
	&=
	\sum_{i=0}^{2^N-1}|a_i(0)|^2e^{-E_i(\tau+\tau^\prime)},
\end{align}
where
$a_i(0)=\langle \phi (0) |(|\phi_i \rangle \langle \phi_i |\otimes \hat{I}_{\B} )|\phi (0) \rangle $ for $i=0,1,\cdots, 2^N-1$.
 By using Eq.\eqref{Eq:expectation basis}, the overlap can be rewritten as
\begin{align}
	&D(\tau,\tau^\prime)=\frac{\braket{\psi(0)|e^{-\hH(\tau+\tau^\prime)}|\psi(0)}}{A(\tau)A(\tau^\prime)}\nn[10pt]
	&=\frac{e^{-2E_0(\tau+\tau^\prime)}}{A(\tau)A(\tau^\prime)}
	\Biggl|\Biggl(m|a_0(0)|^2+\nn
	&\hspace{50pt}\sum_{i= m}^{2^N-1}|a_i(0)|^2e^{-(E_i-E_0)(\tau+\tau^\prime)}\Biggr)\Biggr|^2.
\end{align}
Here, $m$ denotes the degeneracy of the ground state of the Hamiltonian $\hH$ of the system A, and we have
$a_0(0)=a_1(0)=\cdots =a_{m-1}(0)$. For $\tau'_{\infty}\gg \tau^\prime$, we have
\begin{equation}
    m|a_0(0)|^2 \gg |(\sum_{i= m}^{2^N-1}|a_i(0)|^2e^{-(E_i-E_0)(\tau+\tau'_{\infty})})|^2
\end{equation}
 and thus we obtain
 \begin{align}\label{eq:overlapinf}
	D(\tau,\tau'_{\infty})
	\simeq 
	\frac{m^2}{A(\tau)A(\tau'_{\infty})}e^{-2E_0(\tau+\tau'_{\infty})}|a_0(0)|^4.
\end{align}
The value of $\tau'_{\infty}$ can be determined
as follows.  We calculate the energy of the state against $\tau$ during the imaginary time evolution, and define $\tau'_{\infty}$ as a time when the energy almost
converges to a specific value.
We will explain this later when performing numerical simulations.
Now, substituting $\tau =\tau'_{\infty}$ into Eq. \eqref{eq:overlapinf}, we obtain 
\begin{align}
	1=D(\tau'_{\infty},\tau'_{\infty})
	&\simeq 
	\frac{m^2}{A^2(\tau'_{\infty})}e^{-4E_0\tau'_{\infty}}|a_0(0)|^4,
\end{align}
which is rewritten as
\begin{align}\label{Andtau}
	A(\tau'_{\infty})
	&\simeq 
	m e^{-2E_0\tau'_{\infty}}|a_0(0)|^2.
\end{align} 
By combining Eq.~\eqref{Andtau} with Eq. \eqref{eq:overlapinf}, we obtain
\begin{align}
	A(\tau)
	&\simeq 
	\frac{m}{D(\tau,\tau'_{\infty})}e^{-2E_0\tau}|a_0(0)|^2.
\end{align}
Furthermore, using $|a_0(0)|^2=1/2^N$ and $Z(2\tau)=2^NA(\tau)$, we finally obtain
\begin{align}\label{Eq:ROM partition function}
	Z(2\tau)
	&\simeq 
	\frac{m}{D(\tau,\tau'_{\infty})}e^{-2E_0\tau}.
\end{align}
Equation \eqref{Eq:ROM partition function} is the partition function of the Hamiltonian $\hH$. We call this method the Reversed Overlap Method (ROM). It is worth mentioning 
that, 
as the temperature increases,
the overlap $D(\tau,\tau'_{\infty})$
for the ROM becomes smaller, and the necessary number of the measurements
to find the value of $D(\tau,\tau'_{\infty})$
increases.

We can calculate the degeneracy $m$ of the ground state as follows.
After implementing the imaginary time evolution, and tracing out the system B, we have the state 
$\hrho(\tau'_\infty)=\tr_{\B}[\ket{\psi(\tau'_{\infty})}\bra{\psi(\tau'_{\infty})}]$.
We then make use of the relation $m=1/\tr[\hrho(\tau'_\infty)^2]$ between the degeneracy $m$ and the purity $\tr[\hrho(\tau'_\infty)^2]$. Since the purity can be calculated with a method called a destructive SWAP test (see Appendix \ref{DStest} for the details) by preparing two copies of the state, we can determine the degeneracy $m$.



\begin{figure}[h]
		\centering
		\includegraphics[width=\columnwidth]{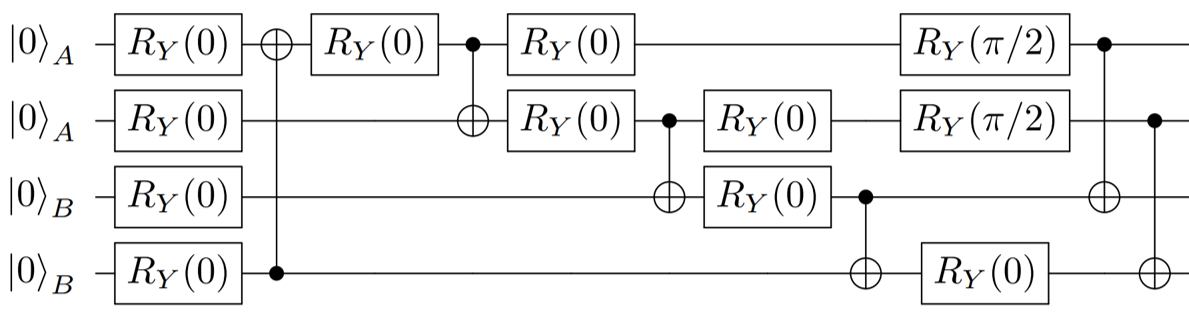}
		\caption{{Ansatz circuit for the variational imaginary time simulation with 4 qubits. $\ket{0}_{\rm{A}}$ and $\ket{0}_{\rm{B}}$ denotes the state with 
		the system A and the system B, respectively. In the description of the rotation operator $R_Y$, we represent the initial parameter $\vec\theta(0)$ for the variational algorithm.}
	}
		\label{fig:ansatz}
\end{figure}

\section{Numerical simulations}
We evaluate the performance of our method by using numerical simulations.
We adopt the Heisenberg model as
\begin{equation}\label{Eq:hshamil}
    \hH=-\sum_{i=1}^{N-1}J
    (\hsig_{x}^{(i)}\hsig_{x}^{(i+1)}+\hsig_{y}^{(i)}\hsig_{y}^{(i+1)}+\hsig_{z}^{(i)}\hsig_{z}^{(i+1)}),
\end{equation}
where $\hsig_{\alpha}^{(k)}\ (\alpha=x,y,z)$ is the Pauli operator acting on the $k$-th qubit, $J$ is the coupling strength between qubits, and $N$ is the number of qubits. 
We perform numerical simulations for the case of $N=2$.
We consider two cases: (i) the coupling strength $J$ is positive or (ii) negative.
For
$J>0$, the ground states are three-fold degenerate:$\ket{00}$, $\frac{1}{\sqrt{2}}(\ket{01}+\ket{10})$, $\ket{11}$.
On the other hand, for $J<0$, the ground state is $\frac{1}{\sqrt{2}}(\ket{01}-\ket{10})$ with no degeneracy.
In Fig.\ref{fig:ansatz},
we show the ansatz circuit for the variational imaginary time simulation and the initial parameters $\vec\theta(0)$.

\begin{figure}[h]
\centering
		\includegraphics[width=\columnwidth]{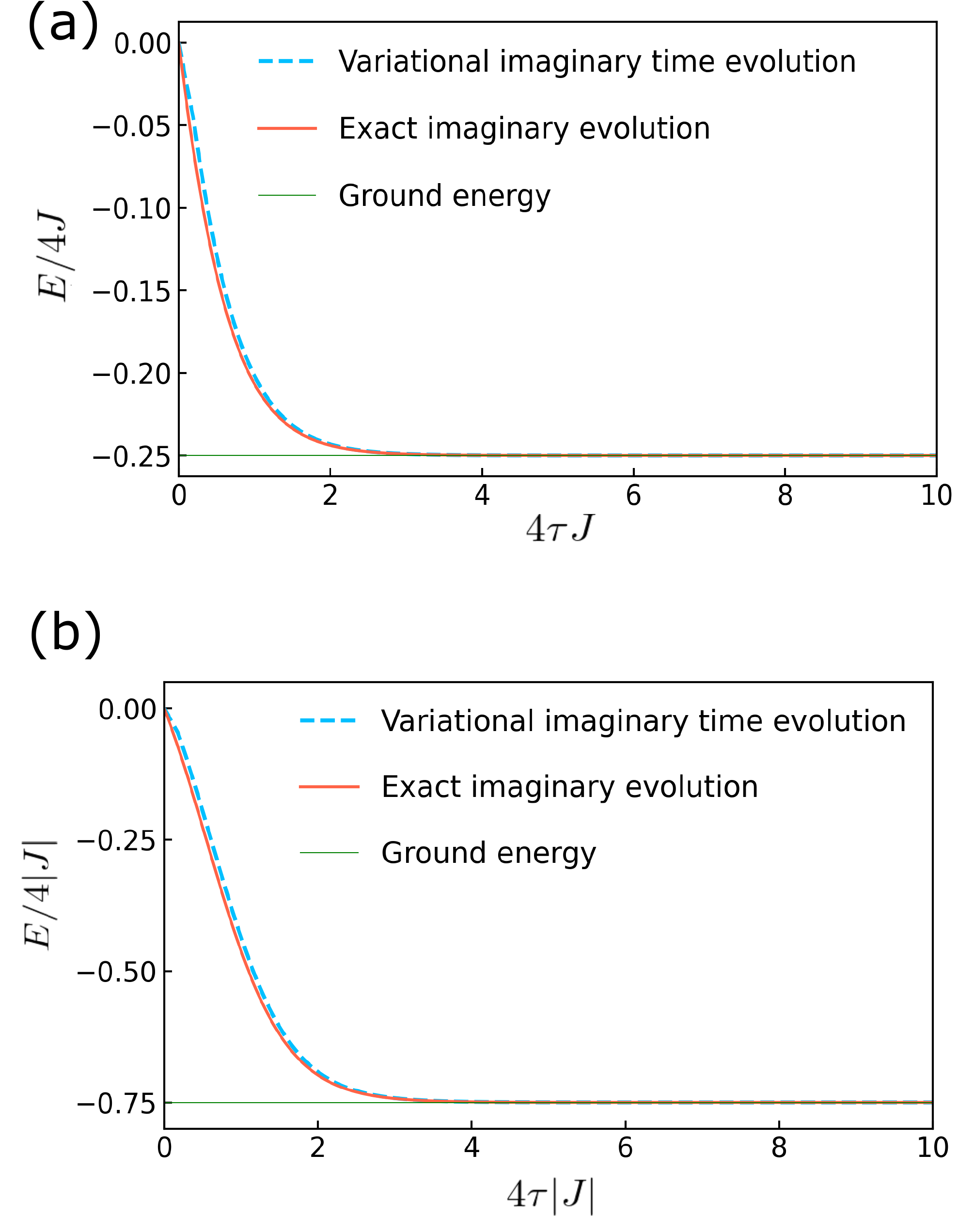} \\
\caption{
Energy of the state during the variational
imaginary time evolution where we adopt the Heisenberg model as the Hamiltonian. 
The dashed red line is the exact imaginary time
evolution, and the blue line is the variational imaginary time
evolution. The green solid line is the ground state energy. 
(a)The coupling constant is set as $J>0$, and the ground state is degenerate. (b)The coupling constant is set as $J<0$, and the ground state is not degenrate.
}
\label{fig:energy}
\end{figure}

We plot the energy (the expectation value of the Hamiltonian) during the imaginary time evolution in Fig. \ref{fig:energy}. 
We confirm that the energy converges to a constant value for a large $\tau$.
Since the energy becomes almost steady around $\tau J=2.5$, we choose $\tau'_{\infty}=2.5/J$.

\begin{figure}[h]
\centering
		\includegraphics[width=\columnwidth]{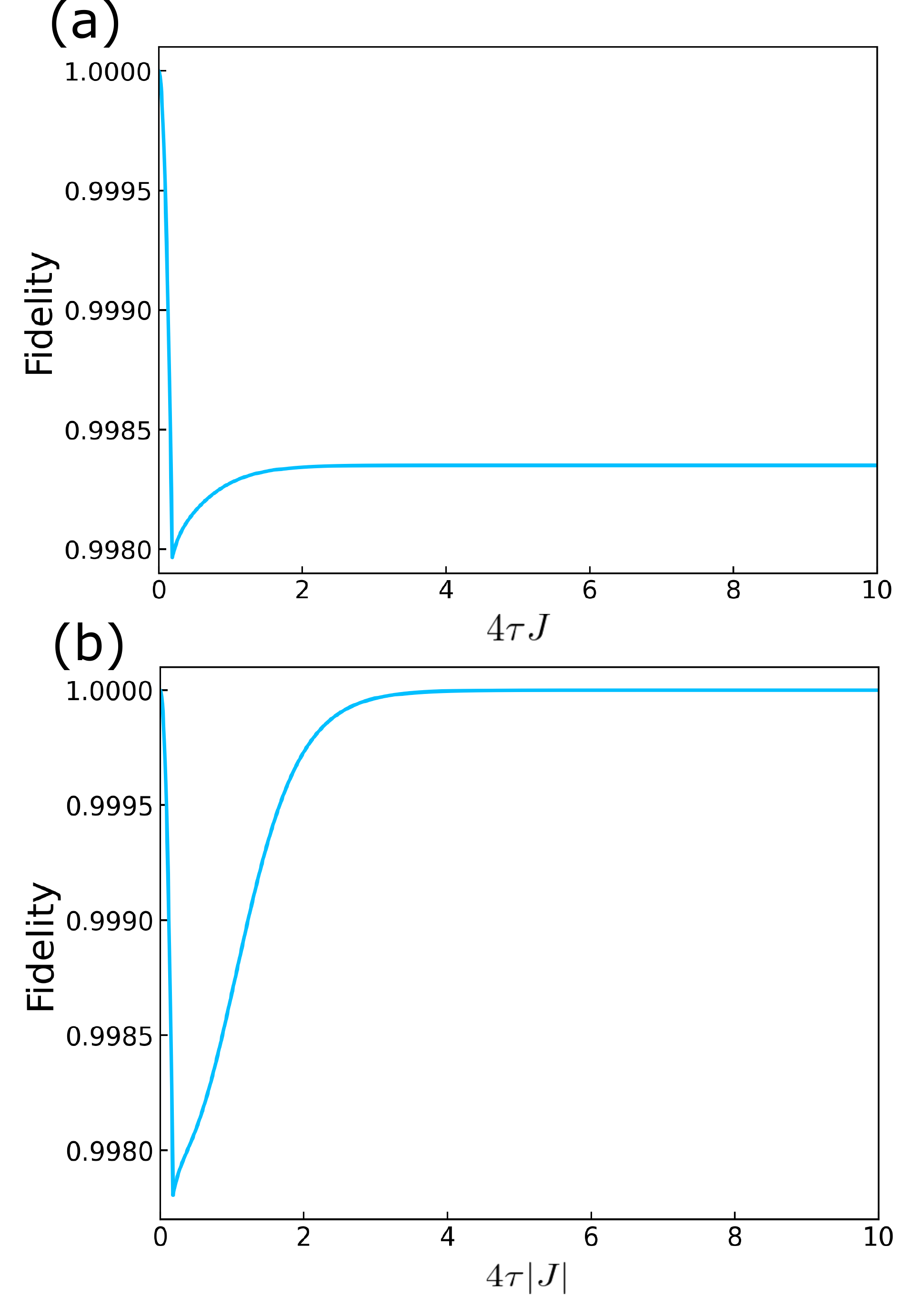} \\
\caption{
Fidelity $F(\ket{\phi(\vec{\theta}(\tau))},\ket{\psi(\tau)})=|\braket{\phi(\vec{\theta}(\tau))|\psi(\tau)}|^2$ between the parametrized wavefunction $\ket{\phi(\vec{\theta}(\tau))}$ and the exact solution $\ket{\psi(\tau)}$ against the time $\tau$ for the imaginary time evolution.
 We adopt the Heisenberg model, and 
 the coupling strength are chosen to be (a) $J>0$ and (b) $J<0$.
}
\label{fig:fidelity}
\end{figure}

In Fig.~\ref{fig:fidelity},we plot the fidelity $F(\ket{\phi(\vec{\theta}(\tau))},\ket{\psi(\tau)})=|\braket{\phi(\vec{\theta}(\tau))|\psi(\tau)}|^2$ between the parametrized wavefunction $\ket{\phi(\vec{\theta}(\tau))}$ for the variatioal imaginary evolution
and the exact state $\ket{\psi(\tau)}$ obtained by solving Eq.~\eqref{imgschsol}.
This shows that our variational quantum circuit accurately simulates the imaginary time evolution.


\begin{figure}[h]
\centering
		\includegraphics[width=\columnwidth]{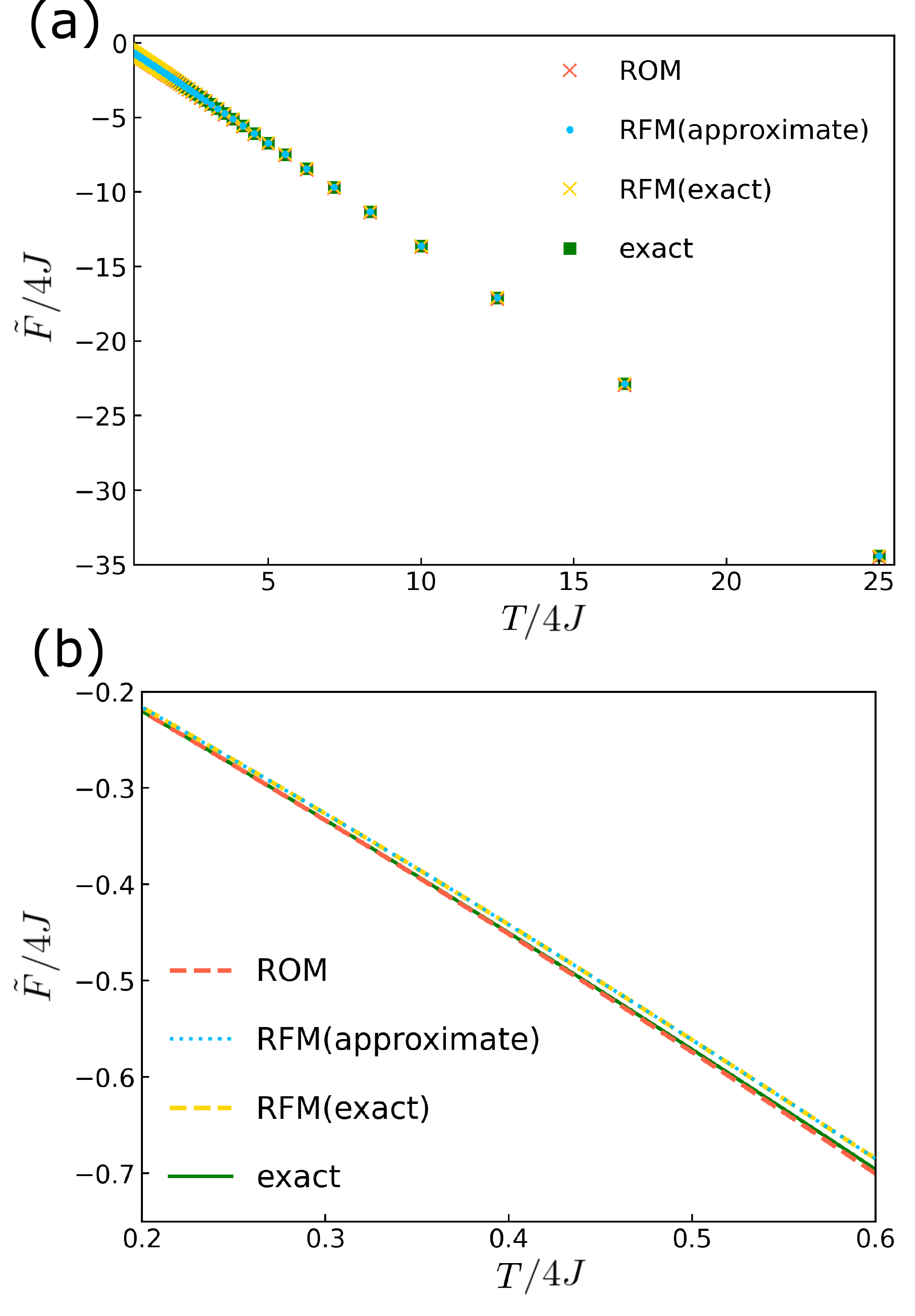} \\
\caption{
The free energy calculated by the RFM and ROM against the temperature. 
In the RFM, we need to calculate the initial value of the partition function at a high temperature by using a classical computer. 
The red plots are the numerical results by the ROM, the blue plots are the numerical results by RFM when we perform the Taylor expansion up to the fourth order to calculate the initial value, the yellow plot are the numerical results by RFM when we use the exact value of the initial value, and the solid green line shows the exact value of the free energy. 
We adopt the Heisenberg model with a positive coupling strength.
}
\label{fig:freeenergy1}
\end{figure}

\begin{figure}[h]
\centering
		\includegraphics[width=8.5cm
		]{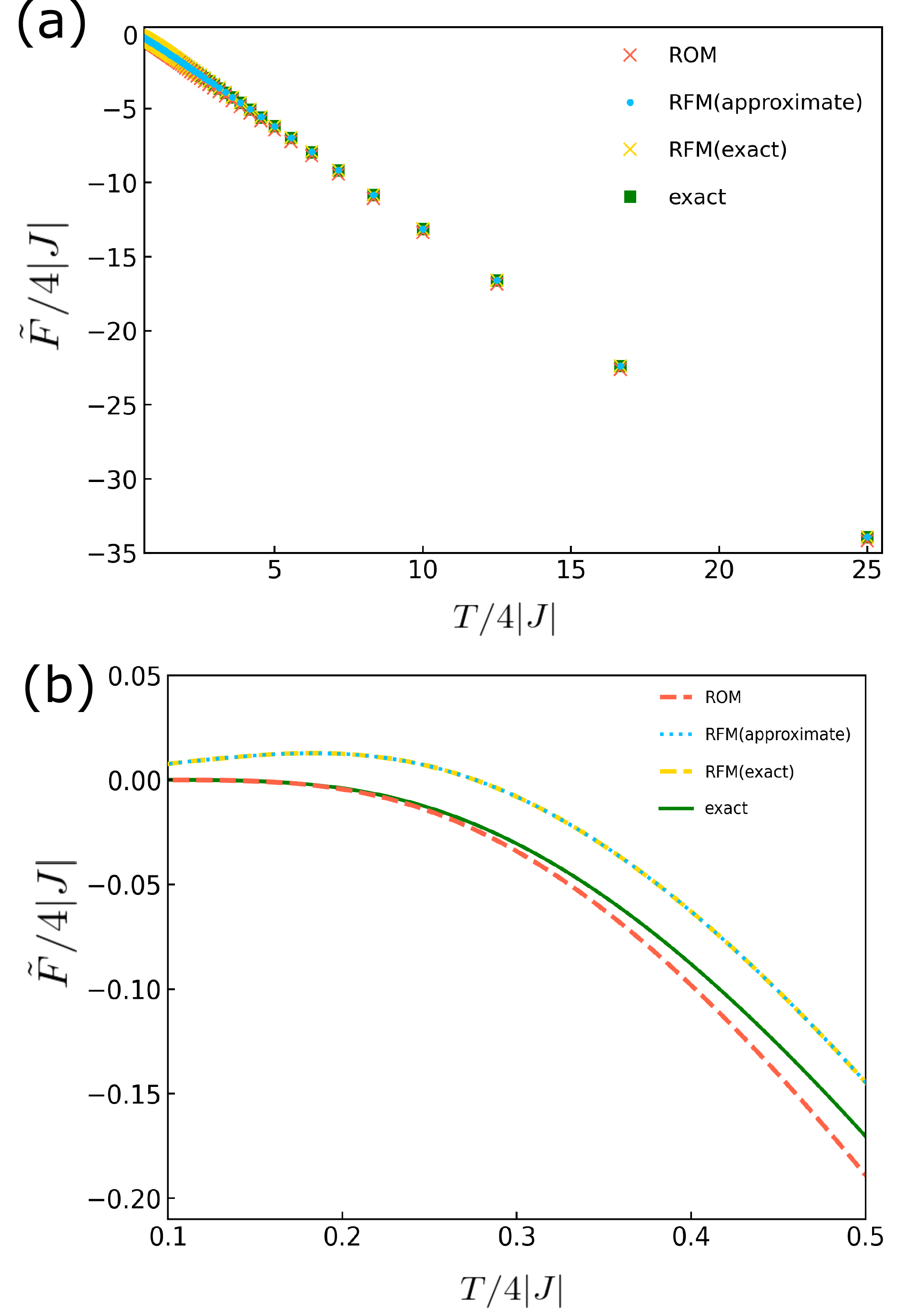} \\
\caption{
The free energy calculated by our method  against the temperature.
We use the same notations as that with the Fig \ref{fig:freeenergy1}. 
We adopt the Heisenberg model with a negative coupling strength.
}
\label{fig:freeenergy2}
\end{figure}

In Figs.~\ref{fig:freeenergy1} and \ref{fig:freeenergy2},
we plot the free energy 
calculated by our method. In Fig.~\ref{fig:freeenergy1}(\ref{fig:freeenergy2}) we show the results for the Heisenberg model with $J>0$ ($J<0$).
There is a good agreement between the exact values and the values calculated with our methods. As we described in Sec.~\ref{theory}, when we calculate the partition function with the RFM, the error tends to accumulate especially at lower temperatuers. On the other hand, ROM does not have such a limitation for low temperatures.
Actually, from Fig.~\ref{fig:freeenergy1} and Fig.~\ref{fig:freeenergy2},
we confirm that the free energy calculated with the ROM becomes closer to the exact value than that with the RFM at low temperatures.


\section{Conclusion}
In conclusion, we propose a scheme to calculate the partition function
by using the variational imaginary time evolution on a near-term quantum computer.
More concretely, we find a systematic way to construct the partition function from the overlap of quantum states during the imaginary time evolution, which does not rely on the extrapolation using R{\'e}ny entropy.
Moreover, the necessary number of the qubits is $2N$ to calculate the partition function of $N$ qubits, which is much smaller than that of the previous approaches. 
Our results show a potential for a practical use of the NISQ device for condensed matter physics.
{\it{Note added.}}\textemdash While preparing our manuscript, we became
aware of a related work that also proposes a scheme to calculate the partition function on a near-term quantum computer \cite{estgibbs2021quantum}.

This work was supported by Leading Initiative for
Excellent Young Researchers MEXT Japan and JST
presto (Grant No. JPMJPR1919) Japan.   S.W. was supported by Nanotech CUPAL,
National Institute of Advanced Industrial Science and Technology (AIST).
This paper was partly based on results obtained from a project, JPNP16007, commissioned by the New Energy and Industrial Technology Development Organization (NEDO), Japan.

\appendix
\section{Destructive SWAP test}\label{DStest}
\begin{figure}[h]
		\centering
		\includegraphics[width=7cm
		]{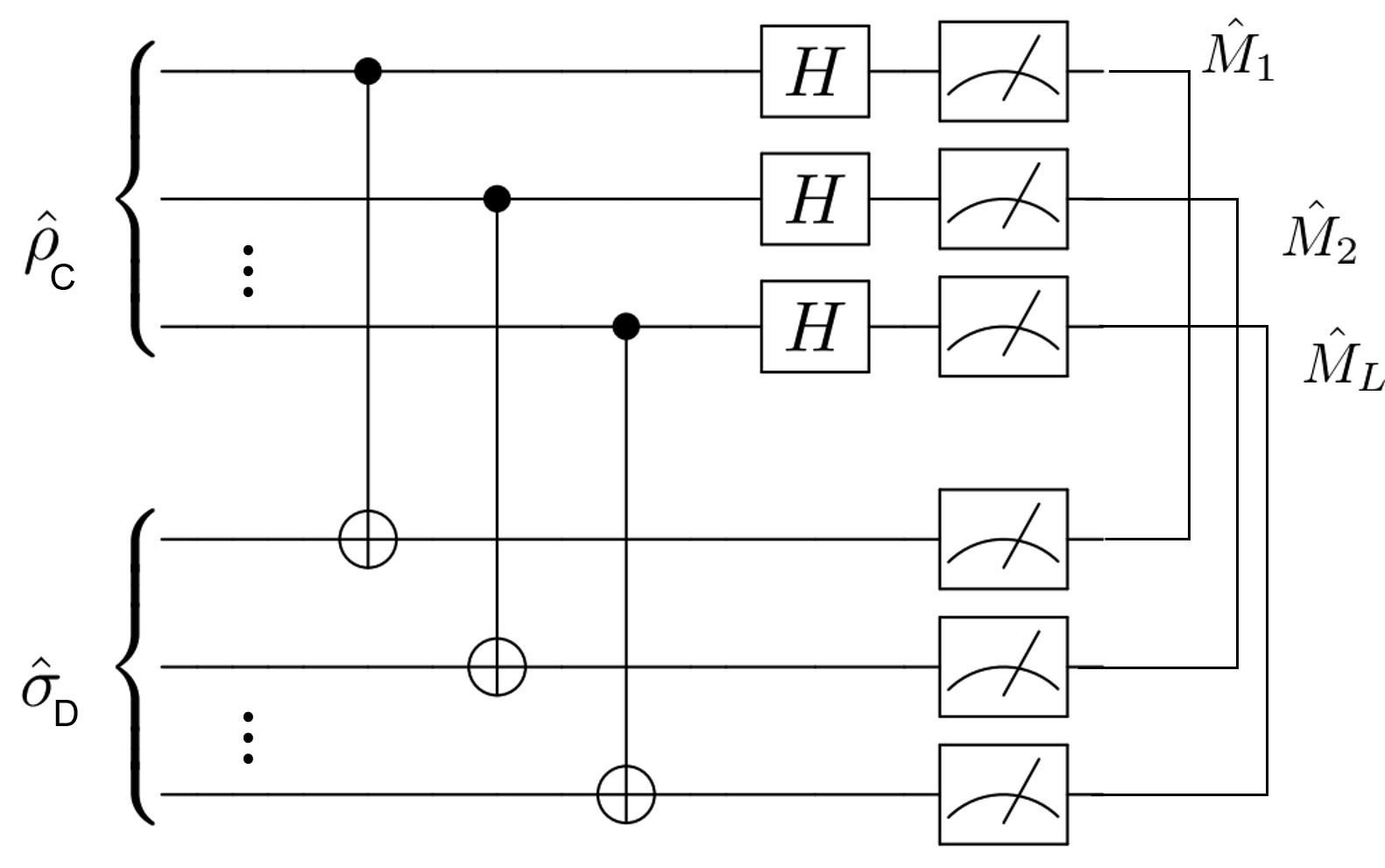} \\
		\caption{A quantum circuit to perform destructive SWAP.
	}
		\label{fig:swap}
	\end{figure}
In this appendix, we review 
a destructive SWAP test.
 This lets us compute the purity of a density matrix if we prepare two copies of the state~\cite{ekert2002direct,garcia2013swap,cincio2018learning}.
We prepare a system C that is composed of $L$ qubits, and prepare the other system D that is also composed of $L$ qubits.
The destructive SWAP test is an algorithm to measure 
$\tr[\hrho_{\C }\hsig_{\C}]$ by using two density matrices $\hrho_{\C}$ and $\hsig_{\D}$.
Here, we assume that
$\hsig_{\C}$ has the same form as $\hsig_{\D}$, but $\hsig_{\C}$ ($\hsig_{\D}$)
is the state in the system C (D). 
We have the relation
\begin{equation}
    \tr\left[(\bigotimes_{n=1}^{L}\hat{M_n})(\hrho_{\C}\otimes\hsig_{\D})\right]=\tr[\hrho_{\C}\hsig_{\C}],
\end{equation}
where
the observable $\hat{M}_n$ is an operator that non-locally acts on the $n$-th qubit of system C and the $n$-th qubit of system D, and can be expressed as follows:
\begin{equation}
    \hat{M}_n=\hat{P}^{(n)}_1+\hat{P}^{(n)}_2+\hat{P}^{(n)}_3-\hat{P}^{(n)}_4.
\end{equation}
Here, $\{\hat{P}^{(n)}_i\}_{i=1,2,3,4}$ represents the projection operator onto the Bell basis
\begin{align}
    \hat{P}^{(n)}_1&=\ket{\psi^{+}}\bra{\psi^{+}},\ \hat{P}^{(n)}_2=\ket{\psi^{-}}\bra{\psi^{-}},\nn[10pt]
    \hat{P}^{(n)}_3&=\ket{\phi^{+}}\bra{\phi^{+}},\ \hat{P}^{(n)}_4=\ket{\phi^{-}}\bra{\phi^{-}}
\end{align}
where the Bell basis is defined as 
\begin{align}
    \ket{\psi^{\pm}}&=\frac{1}{2}(\ket{00}_n\pm\ket{11}_n),\\[5pt]
    \ket{\phi^{\pm}}&=\frac{1}{2}(\ket{01}_n\pm\ket{10}_n),
\end{align}
and $\ket{00}_n$, $\ket{01}_n$, $\ket{10}_n$, $\ket{11}_n$ represents the state of the $n$-th qubits in the systems C and D. The destructive SWAP test can be performed with the quantum circuit shown in Fig.~\ref{fig:swap}. 
It is known that a sequential implementation of a CNOT gate between two qubits, an Hadamard gate on each qubit, and measurements in the computational basis on the two qubits allows us to perform the measurement in the Bell basis. When a projection $\hat{P}^{(n)}_4$ occurs, we assign $-1$ as a measurement result. On the other hand, 
when a projection of either $\hat{P}^{(n)}_1$, $\hat{P}^{(n)}_2$, or $\hat{P}^{(n)}_3$ occurs, we assign $+1$ as a measurement result.
This lets us measure the observable of $\bigotimes_{i=1}^{L}\hat{M_i}$.
By repeating this process and averaging over the measurements, we obtain $\tr[\hrho_{\C}\hsig_{\C}]$.
If we have $\hsig_{\C}=\hrho_{\C}$, we can calculate the purity of the state $\hrho_{\C}$.

\bibliography{matsumoto} %
\end{document}